\documentclass[journal=jacsat,manuscript=article]{achemso}

\usepackage[version=3]{mhchem} 
\usepackage[T1]{fontenc}       
\usepackage{graphicx}					
\usepackage{amsmath}
\setkeys{acs}{usetitle = true}
\usepackage{color}

\title{Electronic Properties of Tetraazaperopyrene Derivatives on Au(111): Energy Level Alignment and Interfacial Band Formation}

\author{Arnulf Stein}
\affiliation{Physikalisch-Chemisches Institut, Universit\"at Heidelberg, Im Neuenheimer Feld 253, 69120 Heidelberg, Germany}
\author{Daniela Rolf}
\affiliation[Freie Universit\"{a}t Berlin] {Fachbereich Physik, Freie Universit\"{a}t
Berlin, Arnimallee 14, D-14195 Berlin, Germany}
\author{Christian Lotze}
\affiliation[Freie Universit\"{a}t Berlin] {Fachbereich Physik, Freie Universit\"{a}t
Berlin, Arnimallee 14, D-14195 Berlin, Germany}
\author{Sascha Feldmann}
\affiliation{Physikalisch-Chemisches Institut, Universit\"at Heidelberg, Im Neuenheimer Feld 253, 69120 Heidelberg, Germany}
\author{David Gerbert}
\affiliation{Physikalisch-Chemisches Institut, Universit\"at Heidelberg, Im Neuenheimer Feld 253, 69120 Heidelberg, Germany}
\author{Benjamin G\"{u}nther}
\affiliation{Anorganisch-Chemisches Institut, Universit\"at Heidelberg, Im Neuenheimer Feld 270, 69120 Heidelberg, Germany}
\author{Andreas Jeindl}
\affiliation{Technische Universit{\"a}t Graz, Institut f{\"u}r Festk{\"o}rperphysik, NAWI Graz, Petersgasse 16, 8010 Graz, Austria}
\author{Johannes J. Cartus}
\affiliation{Technische Universit{\"a}t Graz, Institut f{\"u}r Festk{\"o}rperphysik, NAWI Graz, Petersgasse 16, 8010 Graz, Austria}
\author{Oliver T. Hofmann}
\affiliation{Technische Universit{\"a}t Graz, Institut f{\"u}r Festk{\"o}rperphysik, NAWI Graz, Petersgasse 16, 8010 Graz, Austria}
\author{Lutz H. Gade}
\affiliation{Anorganisch-Chemisches Institut, Universit\"at Heidelberg, Im Neuenheimer Feld 270, 69120 Heidelberg, Germany}
\author{Katharina J. Franke}
\affiliation[Freie Universit\"{a}t Berlin] {Fachbereich Physik, Freie Universit\"{a}t
Berlin, Arnimallee 14, D-14195 Berlin, Germany}
\author{Petra Tegeder}
\affiliation{Physikalisch-Chemisches Institut, Universit\"at Heidelberg, Im Neuenheimer Feld 253, 69120 Heidelberg, Germany}
\email{tegeder@uni-heidelberg.de}
\phone{+49 (0) 6221 548475}
\fax{+49 (0) 6221 456199}

\begin{document}

\begin{abstract}
N-Heteropolycyclic aromatic compounds are promising organic electron-transporting semiconductors for applications in field effect transistors. Here, we investigated the electronic properties of 1,3,8,10-tetraazaperopyrene  derivatives adsorbed on Au(111) using a complementary experimental approach, namely scanning tunneling spectroscopy and two-photon photoemission combined with state-of-the-art density functional calculations.
We find signatures of weak physisorption of the molecular layers, such as the absence of charge transfer, a nearly unperturbed surface state and an intact herringbone reconstruction underneath the molecular layer. Interestingly, molecular states in the energy region of the \emph{sp}- and \emph{d}-bands of the Au(111) substrate exhibit hole-like dispersive character. We ascribe this band character to hybridization with the delocalized states of the substrate. We suggest that such bands, which effectively leave the molecular frontier orbitals largely unperturbed, to be a promising lead for the design of organic-metal interfaces with a low charge injection barrier.

\end{abstract}

\section{Introduction}
 Organic electron-transporting (n-channel) semiconductors are of particular interest for their implementation in field effect transistors \cite{klauk2007}.
Replacing carbon atoms by nitrogen atoms in a $\pi$-conjugated aromatic molecular backbone typically leads to an energetic stabilization of the frontier orbitals, i.e., the electron affinity (EA) and the ionization potential (IP) increase, while the optical gap size is almost uninfluenced \cite{wurthner2011, Bunz2013, Miao2014, Bunz2015}. N-heteropolycyclic aromatic molecules are thus expected to be promising candidates for n-channel semiconductors. Regardless of n- or p-channel semiconducting behavior in a transistor, the energetic positions of the electron affinity level or the ionization potential are the relevant quantities for device performance. The IP can be determined by photoemission spectroscopies (e.g. ultraviolet photoemission spectroscopy (UPS) \cite{Ishii1999, Braun2009, Koch2013, Oehzelt2014}, two-photon photoemission (2PPE) \cite{Muntwiler2010, Bogner2015, Bogner2016, Stein2019}), or scanning tunneling spectroscopy (STS) \cite{Zandvliet2009}. For the identification of affinity levels, inverse photoemission \cite{Hill2010, Hwang2007}, 2PPE or STS can be used. All methods imply that a former unoccupied molecular state is populated with an electron, creating a transient negative ion resonance.
The difference between the IP and EA is the transport gap (IP - EA = E$_\mathrm{transp.}$), which is different from the optical gap (HOMO-LUMO transition leading to exciton formation)\cite{Kahn2016}. The latter can be determined in surface-adsorbed molecules with 2PPE \cite{Muntwiler2010, Bogner2015, Bogner2016, Stein2019, Ajdari2020}, differential reflectance spectroscopy \cite{Gahl2010, Forker2012, Denk2014, Moldt2016}  or high resolution electron energy loss spectroscopy \cite{Bronner2012, bronner2013, navarro2014, Maass2016, Maass2017, Ajdari2020, Ajdari2020a, Ajdari2021}.

\begin{figure}[htb]
\centering
\resizebox{0.4\hsize}{!}{\includegraphics{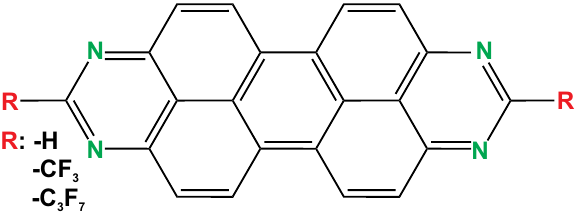}}
\caption{1,3,8,10-tetraazaperopyrene (TAPP) derivatives investigated in the present study.}
\label{molecules}
\end{figure}

The electronic structure of free (gas phase) molecules is strongly affected in the condensed phase and in particular when the molecules are adsorbed on metal surfaces. The adsorption geometry of the organic compound at the hybrid organic/metal interface has a pronounced influence on the interfacial electronic structure \cite{Ishii1999,Koch2013,Gruenewald2013,Braun2009,Oehzelt2014,Mercurio2013,Bredas2009,Koehler2015,May2011,Beljonne2011,Ruhle2011,Avino2016,Casu2015,Forker2012}. The interfacial electronic structure, i.e., the wave function mixing (hybridization) between localized molecular electronic states and metal bands is particularly important for the interfacial charge injection properties and accordingly crucial for the performance of organic field effect transistors. Hybridization is a necessary prerequisite for interfacial band formation which can be identified via dispersing states in angle-resolved photoemission experiments. Interfacial band formation has been shown in the case of strong electron acceptors (tetrafluoro-tetracyanoquinodi-methane (F$_{4}$TCNQ)/Au(111) \cite{Yamane2017, Gerbert2018}, 3,4,9,10-perylene-tetracarboxylic-dianhydride (PTCDA)/ Ag(110) \cite{Wiessner2013} and 1,4,5,8-naphthalenetetracarboxylicdi- anhydride (NTCDA)/Ag(111)  \cite{Wiessner2013, Wiessner2013a}) and a donor (tetrathiafulvalene (TTF)/Au(111) \cite{Gerbert2018}) adsorbed on noble metal surfaces. Recently, we demonstrated interfacial band formation for a N-heteropolycyclic molecule, a 1,3,8,10-tetraazaperopyrene derivative (TAPP-CF$_{3}$, see Fig. \ref{molecules}) adsorbed on Au(111) in the energy regime of occupied and unoccupied electronic states occurred \cite{Stein2021}.
TAPPs belong to the class of N-heteropolycyclic aromatics, which have shown promising results as organic semiconductors in thin film transistors \cite{martens2012, Hahn2015}. The TAPP-H/Cu(111) system has been studied with respect to thermally-activated and surface-assisted reactions using scanning tunneling microscopy (STM) \cite{Matena2008, Matena2010}.

In the present contribution we used complementary experimental techniques, namely STS and 2PPE in combination with density functional theory (DFT) to determine the electronic properties of a 1,3,8,10-tetraazaperopyrene derivative (TAPP-CF$_{3}$, see Fig.\ref{molecules}) adsorbed on Au(111).
The adsorption structure of TAPP-CF$_{3}$ on Au(111) is analyzed by STM and DFT-calculations. 2PPE is applied to study the electronic properties of further TAPP derivatives (TAPP-H and TAPP-C$_{3}$F$_{7}$, see Fig. \ref{molecules}). For all TAPP derivatives we determined the energetic position of several affinity levels as well as the ionization potential. Hybrid band formation in the energy region of occupied and unoccupied delocalized states  are identified \emph{via} angle-resolved 2PPE.

\section{Methods}
STS and 2PPE experiments were carried out with two different setups under ultra-high vacuum (UHV) conditions. In both cases, a clean Au(111) substrate was prepared by a standard procedure
of sputtering-annealing cycles. The TAPP molecules were deposited from an effusion cell held at a temperature of 520 K while the Au(111) surface was kept at room temperature. For the STM experiments the deposited coverage was below one monolayer. This provides clean areas of Au(111) for tip treatment and reference measurements.
To prepare a well-defined monolayer (ML) coverage for the 2PPE experiments a multilayer coverage was deposited and afterwards annealed to 500 K. The desorption of the multilayer was monitored by temperature-programmed desorption measurements (see supporting information).

\textbf{2PPE experiments:}
2PPE measurements enable the quantitative determination of the energetic position of occupied initial and unoccupied intermediate or final electronic states of the adsorbate/substrate system in a pump--probe scheme \cite{Zhu2004, Varene2011, Bronner2012, Bronner2012a, Bronner2013b, Bogner2015, Bogner2016, Tegeder2017, Stein2017}. Detailed photon-energy-dependent measurements are needed to assign the observed photoemission peaks to occupied or unoccupied states. Dispersion relations of the electronic states can be obtained by variation of the detection angle ($\alpha$) as $\hbar k_{\|} = (2 m_{e} \cdot E_{kin})^{1/2} \sin \alpha$,  where $m_{e}$ denotes the free electron mass. $k_{\|}$ = 0 corresponds to electrons detected along the surface normal (i.e., $\alpha$ = 0). The tunable femtosecond laser system, which delivers laser pulses over a wide range of photon energies, and the 2PPE setup are described elsewhere \cite{Tegeder2012}.

\textbf{STM/STS experiments:}
STM and STS measurements have been carried out on a sub-monolayer coverage of TAPP-CF$_{3}$ on Au(111) at a temperature of T = 4.5 K under UHV conditions. The STM tip was prepared by indentation into the clean Au crystal. Differential conductance (dI/dV) spectra and maps were recorded using a lock-in amplifier at a frequency of $f= 937$ Hz with a modulation amplitude of $V_{RMS}$ = 5 mV. The metallic properties of the tip were checked by recording d$I$/d$V$ spectra on clean areas of the Au(111) substrate, where only in the energy regime of the occupied Au(111) band structure the Shockley-type surface state, the sp- and d-bands were detected while the spectrum was featureless otherwise. Additional peaks in the d$I$/d$V$ spectra can thus be associated with molecular resonances.

\textbf{DFT-calculations:}
 State-of-the-art DFT calculations of surface-bound molecules were performed with the FHI-aims package \cite{Blum2009} using the PBE exchange-correlation functional together with van-der-Waals (vdW) corrections based on the TS$^{surf}$ method \cite{Ruiz2012}. FHI-aims employs numeric atom-centered orbitals. For this work, we used the default tight settings for the basis set, except for the onset of the cutoff potential for the gold basis functions, which was increased to 5 {\AA}  to get a more accurate description of the surface dipole. The unit cell was sampled with a $7 \times 7$ generalized Monkhorst-Pack grid and a Gaussian smearing of 0.1 eV. The interface was modelled using periodic boundary conditions, using a 5-layer slab for the metal and approximately 80 {\AA}  of vacuum. To electrostatically decouple the slabs perpendicular to the surface, a dipole correction was employed \cite{Neugebauer1992}. To find the optimal adsorption position, the molecule as well as the two topmost Au layers were allowed to fully relax until the forces fell below 0.02 eV/\AA.

For the analysis of the vibronic nature of sidebands in the d$I$/d$V$ spectra we calculated the  relaxed structures of the neutral and negatively charged free molecule (gas phase), using the B3PW91 functional and the 6-31g(d,p) basis set as implemented in the Gaussian 09 package \cite{Gaussian}.

\section{Results and Discussion}
We first focus on the TAPP-CF$_{3}$/Au(111) system and elucidate the adsorption structure by means of STM and DFT, since it strongly influences the interfacial electronic structure. We then analyze the electronic  properties of a (sub)monolayer coverage of TAPP-CF$_{3}$ molecules on the Au(111) surface in detail using STS and 2PPE, and complement these findings with DFT calculations. Subsequently, a comparison to other TAPP derivatives (TAPP-H and TAPP-C$_{3}$F$_{7}$, see Fig. \ref{molecules}) will be drawn by 2PPE measurements.

\subsection{Adsorption structure of TAPP-CF$_{3}$ on the Au(111)}
\begin{figure}
\centering
\resizebox{0.95\hsize}{!}{\includegraphics{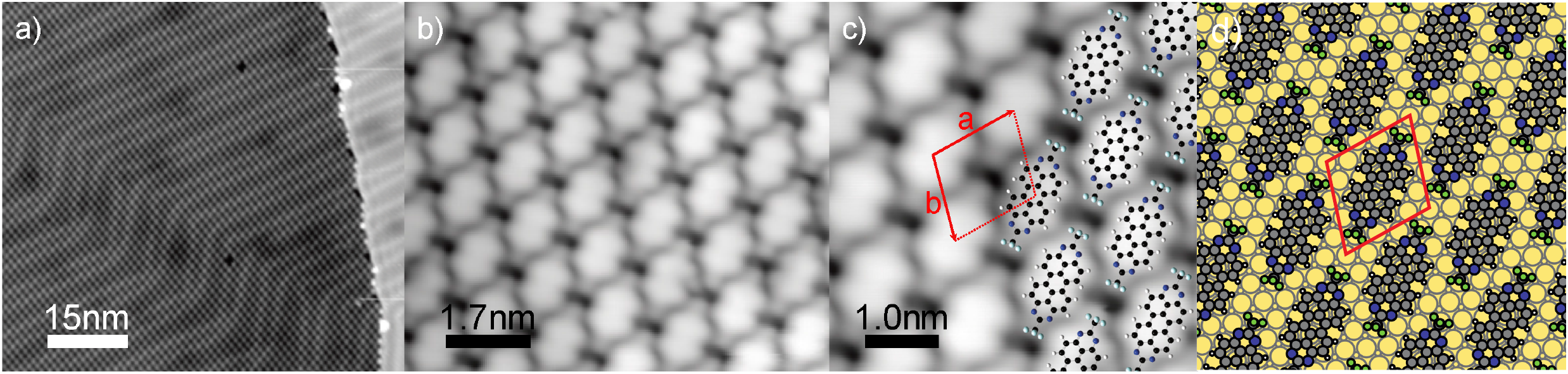}}
\caption{STM topography of a sub-monolayer coverage TAPP-CF$_3$ on Au(111).
a) Large-scale image revealing densely-packed islands with the size limited by monoatomic
steps of the Au(111) surface. $V_{bias}$ = 300 mV, $I_t$ = 300 pA. b) Close-up view of the monolayer showing the individual molecules. $V_{bias}$ = -300 mV, $I_t$ = 300 pA. c) Close-up view with superimposed structure model and unit cell of the molecules periodic lattice. The unit cell size is $a$ = 1.23(5) nm $\times$ $b$ = 1.17(5) nm with an angle of $104(3)^{\circ}$ between $a$ and $b$, resulting in a unit cell area of 1.39(5) nm$^2$. $V_{bias}$ = -300 mV, $I_t$ = 300 pA. d) Theoretical structure model (see text).}
\label{STMtopo}
\end{figure}

STM images reveal that TAPP-CF$_{3}$ molecules form densely-packed extended islands with mono-domain structures up to hundred nanometers diameter (Fig.\,\ref{STMtopo}a). We observe no preferential orientation of the islands with respect to the underlying Au(111) substrate, suggesting a weakly physisorbed state. This scenario is further supported by the herringbone reconstruction remaining intact below the molecular layer.
A close-up view on the molecular arrangements reveals a flat adsorption geometry of the individual molecules (Fig.\,\ref{STMtopo}b), and allows for a structural model of the molecular layer (Fig.\,\ref{STMtopo}c).

To understand the formation of this self-assembled structure, we performed dispersion-corrected DFT calculations. In a first step, we calculated the interaction energy between two parallel aligned TAPP-CF$_{3}$  molecules in the gas phase at different relative positions and distances (see Supporting Information). The energetically beneficial relative positions obtained with this approach qualitatively agree very well to the experimental unit cell (Fig. \ref{STMtopo}d).
 In a second step, we determined the adsorption site of TAPP-CF$_{3}$ by running several full geometry optimizations, with the initial position of the TAPP-CF$_{3}$ molecule on the four different high-symmetry-points of the Au(111) surface (top, bridge, hcp and fcc hollow). The energetically most favorable adsorption site is found when the center of the molecule is above a bridge position. The other adsorption sites are between 10 (top) and 50 (both hollow sites) meV energetically less favorable.
 From this it can be concluded that the monolayer structure is mostly determined by intermolecular interactions (as opposed to molecule-substrate interactions), in line with previous conclusion (see above) that the molecule is mostly physisorbed.
  The adsorption energy (at the bridge site) has an overall value of -2.585 eV,  which can be split into of a vdW-contribution of -2.852 eV and an electronic contribution (+0.262 eV). The latter has a positive value indicating that the adsorption is mainly governed by vdW-forces.

\subsection{Electronic structure of TAPP-CF$_{3}$ on Au(111)}

\begin{figure}
\centering
\resizebox{0.7\hsize}{!}{\includegraphics{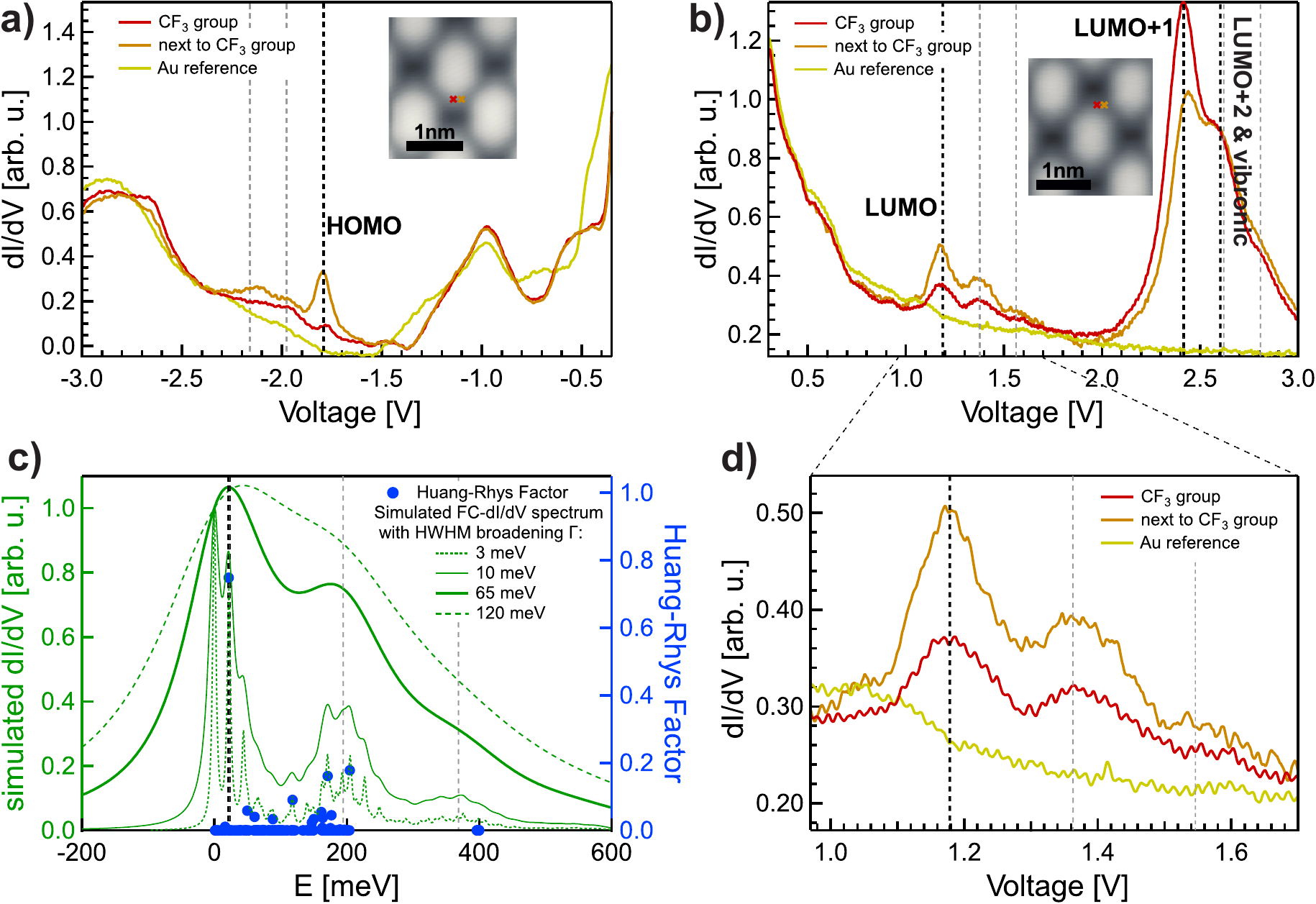}}
\caption{a), b) and d) Tunneling spectra recorded at negative (a) and positive (b) bias voltage at different positions on an individual TAPP-CF$_{3}$ molecule. (d) shows a zoom into the region around the LUMO resonance around 1.0 V to 1.7 V. The background spectrum of the bare Au(111) surface is displayed in yellow. The insets show STM images with color markers to indicate where the individual spectra were taken. The intensity of the molecular resonances varies strongly across the molecule. The feedback parameters for the STM images were $V_{bias}$ = 0.3 V and $I_t$ = 300 pA. The spectra were recorded in constant-current mode with a tunneling current of $I_t$ = 1 nA.
c) Huang-Rhys Factors (blue dots) and simulated Franck-Condon $dI/dV$ spectra (green lines)  of TAPP-CF$_3$ obtained from gas phase DFT simulations (for details see Methods section and Ref. \cite{Krane2018}). Different broadenings were used to simulate different line shapes. A HWHM broadening of 65 meV  produces line shapes that agree well with the experimental line shapes of HOMO, LUMO and LUMO+1.}
\label{STS}
\end{figure}
To resolve the electronic states of the TAPP-CF$_{3}$ molecules, we recorded differential conductance spectra (d$I$/d$V$) on the molecular layer. The spectra of the occupied (negative bias voltage) states are shown in Fig.\,\ref{STS}a (red and orange spectra) and compared to a background spectrum on the bare Au(111) (yellow). First, we note that the broad peak at -1.0 eV is also present in the spectrum on the bare surface and can be associated to the $sp$-band of Au. Second, the surface state of the Au(111) substrate is shifted towards the Fermi level. Third, we observe an additional resonance at -1.8 eV, followed by a broader satellite structure. These features are only present on the molecular layer and may be associated to tunneling through occupied molecular derived states.
Spectra in the regime of the unoccupied states reveal two resonances at 1.2 V and 1.4 V, and two resonances around 2.35 V and 2.55 V that only appear on the molecules. Notably  the spectra shown in Fig. \,\ref{STS}a,b are recorded on and in close vicinity of the CF$_{3}$-group and exhibit variations in the relative intensity of the peaks.

\begin{figure}
\centering
\resizebox{0.6\hsize}{!}{\includegraphics{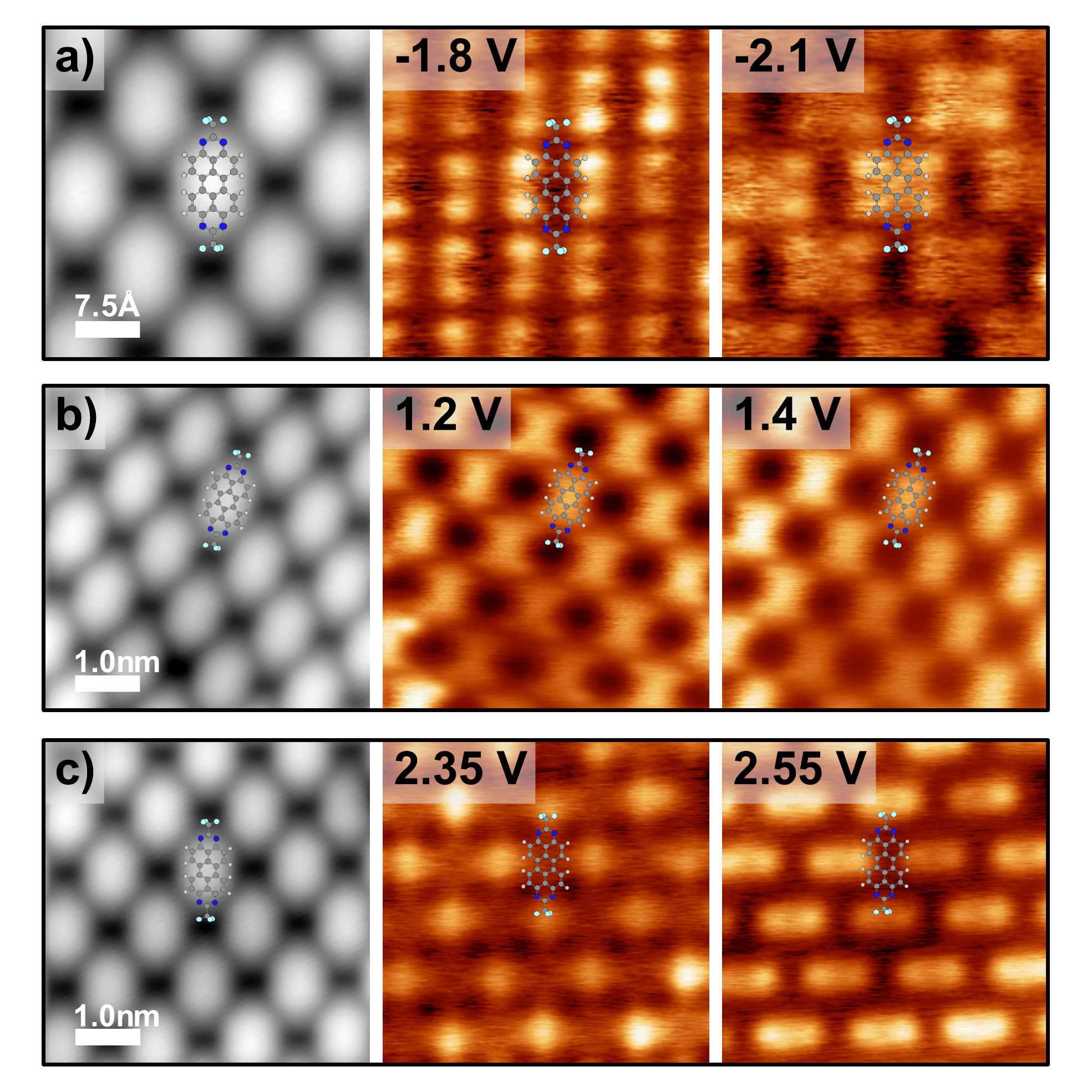}}
\caption{STM images (grey) and differential conductance maps (color) recorded in the same area as the corresponding STM images. The conductance maps were recorded in constant-height mode at the energies of the resonance peak positions, as indicated in the maps. a) Conductance map at the energy of the molecular HOMO. b) Conductance maps at the energy of the LUMO. c) Conductance maps at the energies of the molecular LUMO+1 and LUMO+2. The setpoints for the STM images and before taking the maps were for a) and c) $V_{bias}$ =0.3 V, $I_t$ = 300 pA and for b) $V_{bias}$ =0.5 V, $I_t$ = 300 pA.}
\label{STSmaps}
\end{figure}
A better perception of the variations can be gained from differential conductance maps at the respective energies. These are shown in Fig. \,\ref{STSmaps}. The maps at 1.2 and 1.4 eV show largest intensity all over the TAPP-CF$_{3}$'s backbone, in agreement with the intensity of these peaks decaying quickly away from the molecule. Interestingly, both maps are very similar. In contrast, the maps at 2.35 and 2.55 eV exhibit the largest intensity at the CF$_3$ terminations. Again both maps resemble each other. The distinct spatial intensities can be ascribed to tunneling through different molecular orbitals. Tentatively, we assigned these to the lowest unoccupied molecular orbital (LUMO at 1.2 eV) and LUMO+1 (at 2.35 eV). This assignment will be corroborated by DFT calculations below. The appearance of satellite peaks with the same spatial extent as the main peak is typically ascribed to vibronic states, which arise from the simultaneous excitation of electronic states and vibrational modes in the tunneling process \cite{Qiu2004, Pradhan2005,Nazin2005,Frederiksen2008,Matino2011,Schulz2013,Wickenburg2016}. Further indication of this process is given by the similar spacing of the satellite peaks in the LUMO and LUMO+1.

To sustain the interpretation of the vibronic nature of the sidebands, we calculated the expected d$I$/d$V$ lineshape within the Franck-Condon picture. For this, we considered a free molecule in gas phase and calculated the relaxed structures of the neutral and negatively charged molecule. From these, we derived the vibrational modes and their corresponding Huang-Rhys factor \cite{Krane2018}, which mimics the electron-vibration coupling strength. Fig.\,\ref{STS}c shows the Huang-Rhys factors for all modes. There are several modes with a large Huang Rhys factor, most notably at 22 meV, 50 meV, 118 meV, 162 meV, 172 meV, 177 meV, and 205 meV. To simulate the d$I$/d$V$ spectrum, we considered  individual and coupled modes  with a significant Huang Rhys factor, i.e.,  higher harmonics of individual and coupling of different modes and their harmonics ("progression of progressions"). Then the intensities are convoluted with a Lorentzian lineshape \cite{Krane2018,Yousofnejad2020}.
 A set of spectra with different Lorentzian widths is shown in Fig. \,\ref{STS}c (green).
The best agreement with the experimental lineshape of the LUMO and LUMO+1 (consisting of a broad peak and a shoulder) is obtained when applying a Lorentzian width of 65 mV (half width at half maximum).
 We note that the simulations reveal an additional shoulder at $\sim$400 meV above the main resonance, which can also be found as faint signal in the experimental spectra (see close-up view on the LUMO in Fig. \,\ref{STS}d). The strong similarity of the spectral lineshape of calculated vibronic states and experimental d$I$/d$V$ spectra, together with the similar spatial extent of the main peak and satellite, strongly suggest a vibronic origin of the satellite peak structure. However, we note that fine details in the relative intensity of main resonance and satellite peak vary across the molecule (compare red and orange spectra in Fig. \,\ref{STS}b,c). This behavior  is not captured in the Franck-Condon picture. We suggest that vibration-assisted tunneling can account for these intensity variations \cite{Pavlicek2013,Reecht2020}.

We briefly note that the resonance at -1.8 eV is also followed by a satellite structure, albeit with even less resolution. We suggest that also in this case vibronic peaks contribute to the spectral intensity as the energy spacing matches the one at positive bias voltages.

\begin{figure}
\centering
\resizebox{0.4\hsize}{!}{\includegraphics{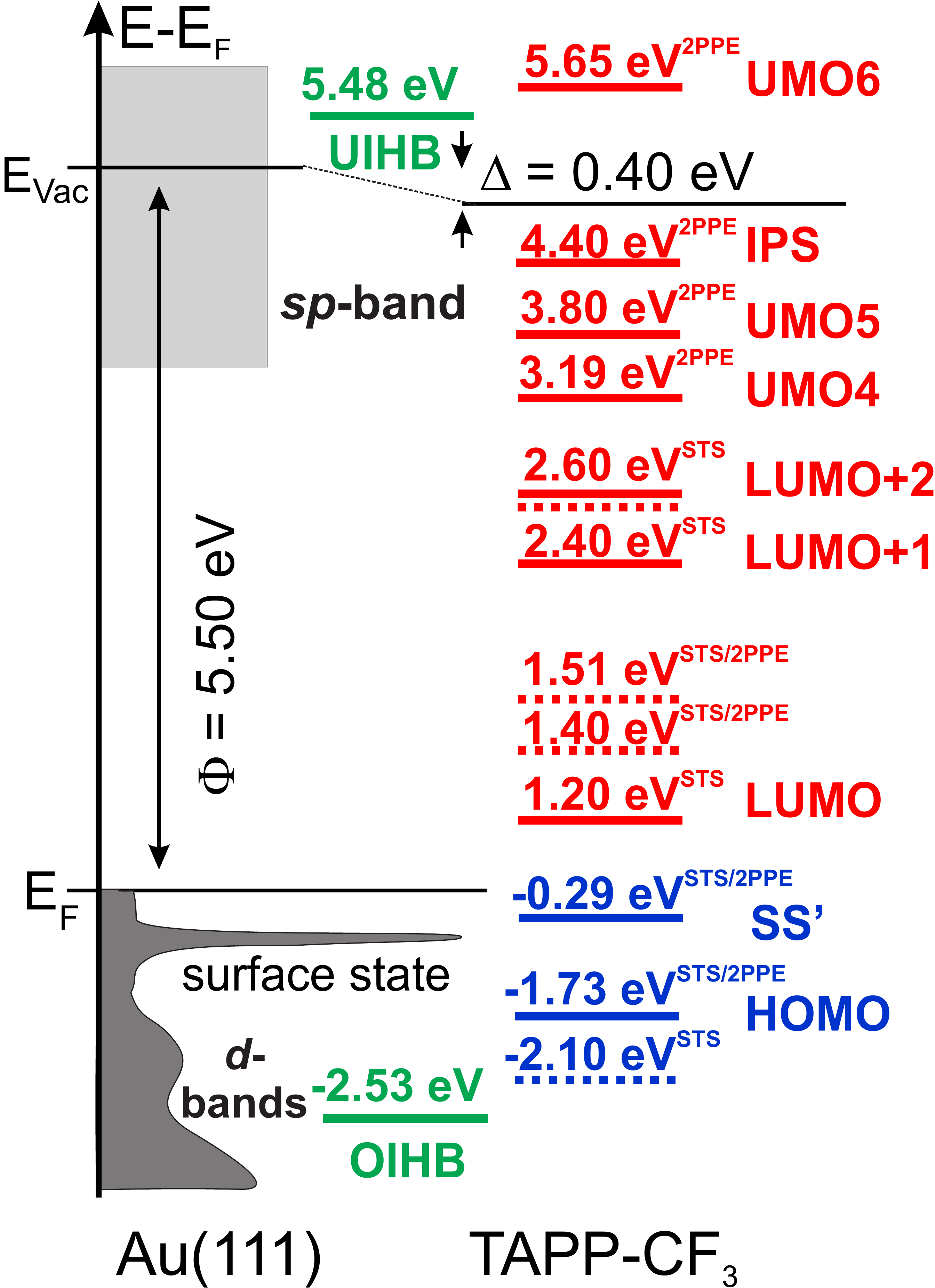}}
\caption{Energy level diagram of  TAPP-CF$_{3}$ adsorbed on Au(111) based on  STS and 2PPE \cite{Stein2021} measurements in the sub- to monolayer regime. Dashed lines indicate vibronic transition. $E_F$ denotes the Au(111) Fermi level and $\Phi$ the work function. UMO refers to an unoccupied molecular orbital, OIHB denotes an occupied interfacial hybrid band and UIHB names an unoccupied interfacial hybrid band. $SS^{\prime}$ refers to the shifted surface state and IPS denotes an image potential state.}
\label{energydiagram_CF3}
\end{figure}

Complementary insight into the electronic structure of TAPP-CF$_{3}$ molecules on Au(111) can be gained by 2PPE measurements. A detailed analysis of the observed peaks and their  photon energy dependency has been published in Ref.\cite{Stein2021}. Here, we compile these results together with the data from the tunneling spectra in Fig.\,\ref{energydiagram_CF3}.
We find very good agreement between the two experimental techniques.
In summary, deposition of 1 ML TAPP-CF$_{3}$ leads to a work function ($\Phi$) decrease of 0.4 eV compared to the bare Au(111)
surface ($\Phi$ = 5.50 eV). Again, we find that the LUMO of TAPP-CF$_{3}$ is energetically far above $E_{F}$ (1.20 eV) and the HOMO far below $E_{F}$ (-1.73 eV).  Thus no Fermi-level pinning occurs and accordingly no charge transfer takes place. Both the LUMO (electron affinity level) and HOMO (ionization potential) are transport states, hence the transport gap is 2.93 eV. Notably the vibronic contributions related to the LUMO are also observed in 2PPE. The higher lying affinity level, the LUMO+1 (2.35 eV) is not seen in 2PPE. This might be due to a weak wave function overlap (transition dipole moment). Two further unoccupied intermediate states are found in 2PPE at 3.19 and 3.80 eV, an energy region in which STS measurements are not feasible. In addition, an unoccupied final state at 5.65 eV is observed. Notably, an occupied interfacial hybrid band (OIHB) and an unoccupied interfacial hybrid band (UIHB) has been identified as discussed in detail in Ref. \cite{Stein2021} (see below).

To corroborate our interpretation of the molecular-derived resonances, we employ DFT calculations to the surface-adsorbed molecule.
First of all, our calculations show that upon adsorption of TAPP-CF$_{3}$, the work function of the interface is reduced by 0.24  eV compared to the pristine Au(111) surface.  This is in excellent agreement with the experiment, which found a work function reduction of 0.40 eV. The reduction in work function can be explained by an induced interface dipole. Computationally, the interface dipole can be separated into a molecular contribution (which is obtained by calculating the potential jump for the hypothetical, free standing monolayer in the geometry it adopts on the surface) and a contribution from the interaction between metal and molecule (often termed "bond dipole"). Here, the molecular contribution amounts to +0.26 eV, arising from the fact that the CF$_{3}$-groups bend out of the molecular plane during adsorption (see supporting information). The bond dipole consequently amounts to -0.50 eV. We note in passing that the computational method we employ here (PBE+TS$^{surf}$) has a tendency for inducing molecular bending upon adsorption \cite{Romaner2007}, thus overestimating the molecular dipole, while underestimating the adsorption distance \cite{Maurer2016, Maurer2019, Hofmann2021}. As a consequence, the effect of Pauli push back is overestimated. The latter describes the effect that the electron density spilling out from the surface is pushed back by the $\pi$ conjugated electron system of the molecular adsorbate layer. As a result, it requires less energy to remove electrons from the substrate and the work function of the system is lowered \cite{Terentjevs2011Interfacial, hofmann2010work}. The slight underestimation of the work function reduction by theory (compared to the experiment) is thus fully consistent with expectations.

\begin{figure}
\centering
\resizebox{0.45\hsize}{!}{\includegraphics{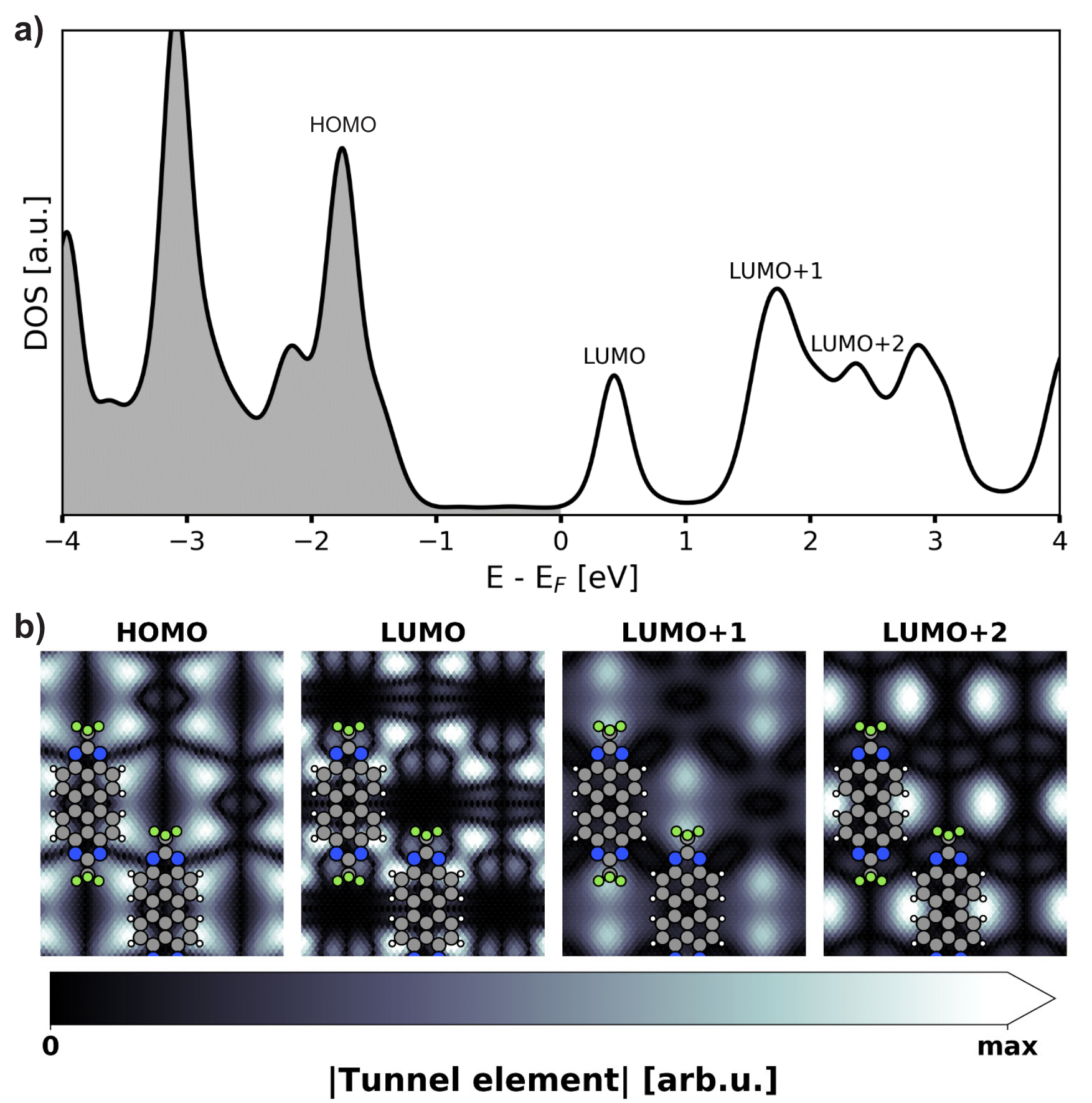}}
\caption{a) Calculated density of states (DOS) projected on the adsorbed molecule. b) Computed tunnel element maps for the orbitals HOMO - LUMO+2. The elements were obtained as the overlap integral between the respective eigenstate of a free-standing molecule and a Au 6s state (representing the tip). The Au state was centered 8 {\AA} above the substrate (corresponding to approx. 3 {\AA}  above the F atoms and <5 {\AA} above the carbon backbone) and moved in the x-y-plane to probe different tip positions.  All maps are normalized to their maximum intensity.}
\label{calculations}
\end{figure}

Remarkably, both the theoretical and the experimental interface dipole values are relatively small. They are smaller than the work function reductions usually found for systems that interact only via Pauli pushback, which for inert organic molecules on gold often amounts to up to 1 eV \cite{Mizushima2017, Koch2007}. An almost zero net interface dipole has been observed for the low-coverage phase of hexaazatriphenylene-hexanitrile (HATCN) on Ag(111) \cite{Glowatzki2008}. However, in contrast to HATCN/Ag(111), for TAPP-CF$_{3}$/Au(111) we find that the molecule is $\approx$ 0.2e positively charged (when applying the Mulliken charge partitioning scheme, see supporting information), while the LUMO remains empty (see Figure \ref{calculations}a).
We therefore explain the relatively small interface dipole to the fact that the bulky CF$_{3}$-group slightly lifts the molecule from the surface, resulting in a relatively large adsorption height and, correspondingly, to a small dipole  as the push back is reduced. This hypothesis is also consistent with the observation that the TAPP-H derivative shows a larger (more negative) and the C$_{3}$F$_{7}$-derivative shows a smaller (less negative) interface dipole experimentally (see below).

Inspection of the density of states (DOS) projected onto the adsorbed molecule (Fig. \ref{calculations}a) reveals that the LUMO is located close to the Fermi-energy (at +0.43 eV). This may easily be mistaken as an indication for Fermi-level pinning, which is not the case. The LUMO peak is more than 3 times the broadening value (0.1 eV) from the Fermi-energy. Hence, the LUMO does not become occupied, which would be a prerequisite for Fermi-level pining. Rather, the proximity of LUMO and Fermi-energy is coincidental, and partly due to inherent shortcomings of the DFT methodology, as we discuss below. Besides the LUMO at +0.43 eV, our projected DOS shows strong molecular peaks at +1.73 eV and +2.36 eV, originating from the LUMO+1 and the LUMO+2, respectively. The DOS is thus qualitatively in agreement with the observed experimental values (e.g. the LUMO/LUMO+1 separation of 1.3 eV), although the energetic position of the states is consistently at too low energies. That there is no perfect agreement between theory and experiment is, per se, not surprising.
Technically, Kohn-Sham orbitals are not physical observables, but they can be associated to ionization energies under the right circumstances \cite{Koerzdoerfer2009}. More importantly, DFT consistently underestimates the energies of empty states, but it also misses the stabilization of ions near a surface through image-charge effects \cite{Biller2011}. At metal/organic interfaces, the latter two effects often cancel each other out, although not perfectly. There are elaborate schemes to achieve better qualitative agreement (e.g., via the use of hybrid functionals and image charge correction schemes \cite{Egger2015}), but these are computationally extremely expensive.

Instead, in order to verify the experimental assignment, we computed the expected dI/dV maps for the HOMO and the first three unoccupied states. To this end, we calculated the overlap integrals between a hypothetical Au-tip, consisting of a 6s orbital, and the eigenstate densities of the molecular monolayer (after removing the substrate for computational effort). In Figure \ref{calculations}b we display maps of these tunnel elements for the HOMO and for LUMO - LUMO+2 when scanning the tip across the surface at a height of 8 {\AA} above the substrate (3 {\AA} above the fluorine atoms). The map of the HOMO agrees well with the map at -1.8V, thus corroborating its previous assignment in experiment (see Fig. \ref{STSmaps}). The theoretical map of the LUMO exhibits a rich structure of nodal planes along the molecular backbone. In experiment, we cannot resolve these details, but the overall intensities at the molecules' center in the dI/dV maps at 1.2 V and 1.4 V is consistent with the interpretation as being LUMO derived. The LUMO+1 has the largest tunnel matrix element at the CF$_{3}$-group, similar to the experimental maps at 2.35 V. As the LUMO+2 is found 200 meV above the LUMO+1, we suggest that it overlaps with the vibronic peaks of the LUMO+1. The map at 2.55 V is thus most probably a convolution of vibronic states and LUMO+2.

\subsection{Electronic structure of further TAPP derivatives on Au(111)}
In the following, we study the influence of the side-chains on the interfacial electronic structure, in particular with respect to interfacial band formation of the other two TAPP derivatives, namely TAPP-H and TAPP-C$_{3}$F$_{7}$ (see Fig. \ref{molecules}) adsorbed on Au(111) by means of 2PPE.

\begin{figure}
\centering
\resizebox{0.4\hsize}{!}{\includegraphics{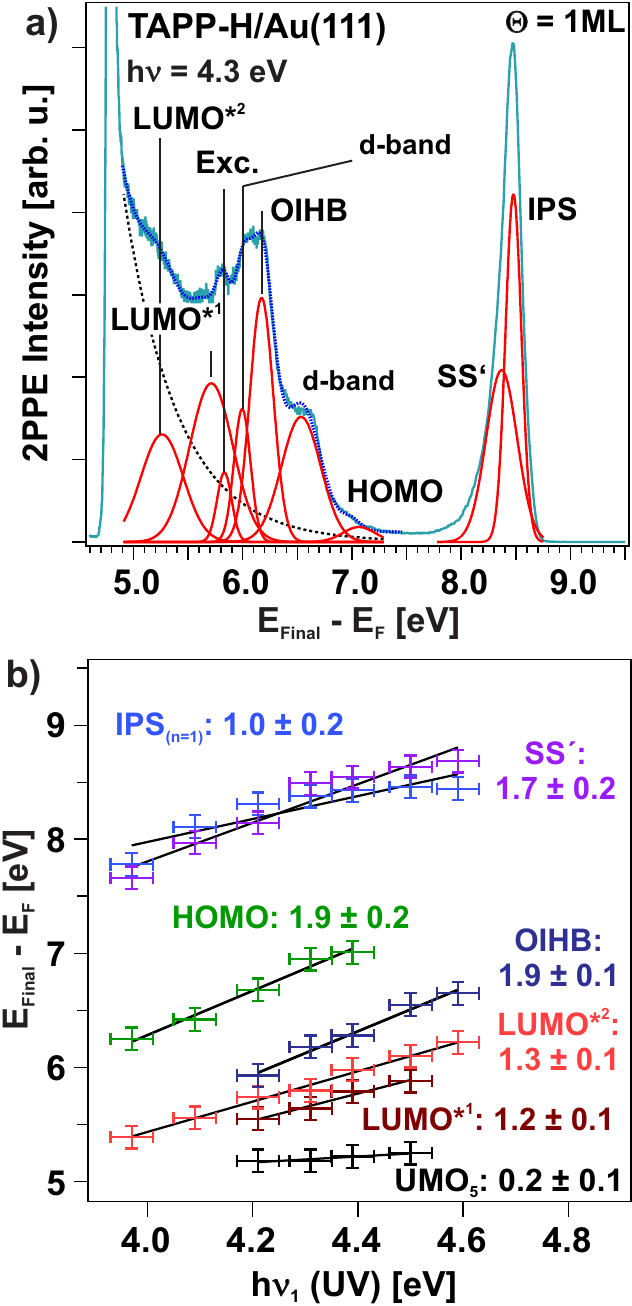}}
\caption{a) 2PPE spectrum recorded with a photon energy of $h\nu$ = 4.3 eV at 1 ML TAPP-H/Au(111).  The energy axis reveals the final state ($E_{Final}$) of photoemitted electrons with respect to the Fermi energy $E_{F}$ ($E_{Final} - E_{F} = E_{kin} + \Phi$); thus, the low-energy cutoff corresponds to the work function ($\Phi$) of the adsorbate/substrate system. The spectrum is fitted by an exponential background and Gaussian-shaped peaks. (b) Photon-energy-dependent peak positions to assign the observed photoemission signals to occupied, unoccupied intermediate or final electronic states. A slope of 1 suggests that a peak originates from an unoccupied intermediate state, a slope of zero from an unoccupied final state (located above the vacuum level), while a slope of 2 is related to peaks originating from occupied states. LUMO* indicate vibronic contributions related to the LUMO.}
\label{2PPE_TAPP_H}
\end{figure}
Figure \ref{2PPE_TAPP_H}a displays a 2PPE spectrum of 1 ML TAPP-H/Au(111) recorded with a photon energy of 4.3 eV. The spectrum is fitted by an exponential background and Gaussian-shaped peaks. Several 2PPE features
are observed. We use the characteristic behaviour of photon-energy-dependent measurements to assign peaks to occupied or unoccupied electronic states (Fig. \ref{2PPE_TAPP_H}b). Additional 2PPE data are presented in the supporting information. Apart from occupied and unoccupied electronic TAPP-H-derived states, the Au(111) d-bands, the shifted surface state (SS$^{\prime}$) as well as the first ($n$ = 1) image potential state
(IPS) are observed. Moreover a peak deriving from an OIHB is identified  (see below).
The observed energy levels are compiled in Fig. \ref{energydiagram}, right column.
An analogous measurement and analysis protocol has been applied to a monolayer of TAPP-C$_{3}$F$_{7}$ on Au(111) (data can be found in the supporting information). The results are also shown in Fig. \ref{energydiagram}.
\begin{figure}
\centering
\resizebox{0.7\hsize}{!}{\includegraphics{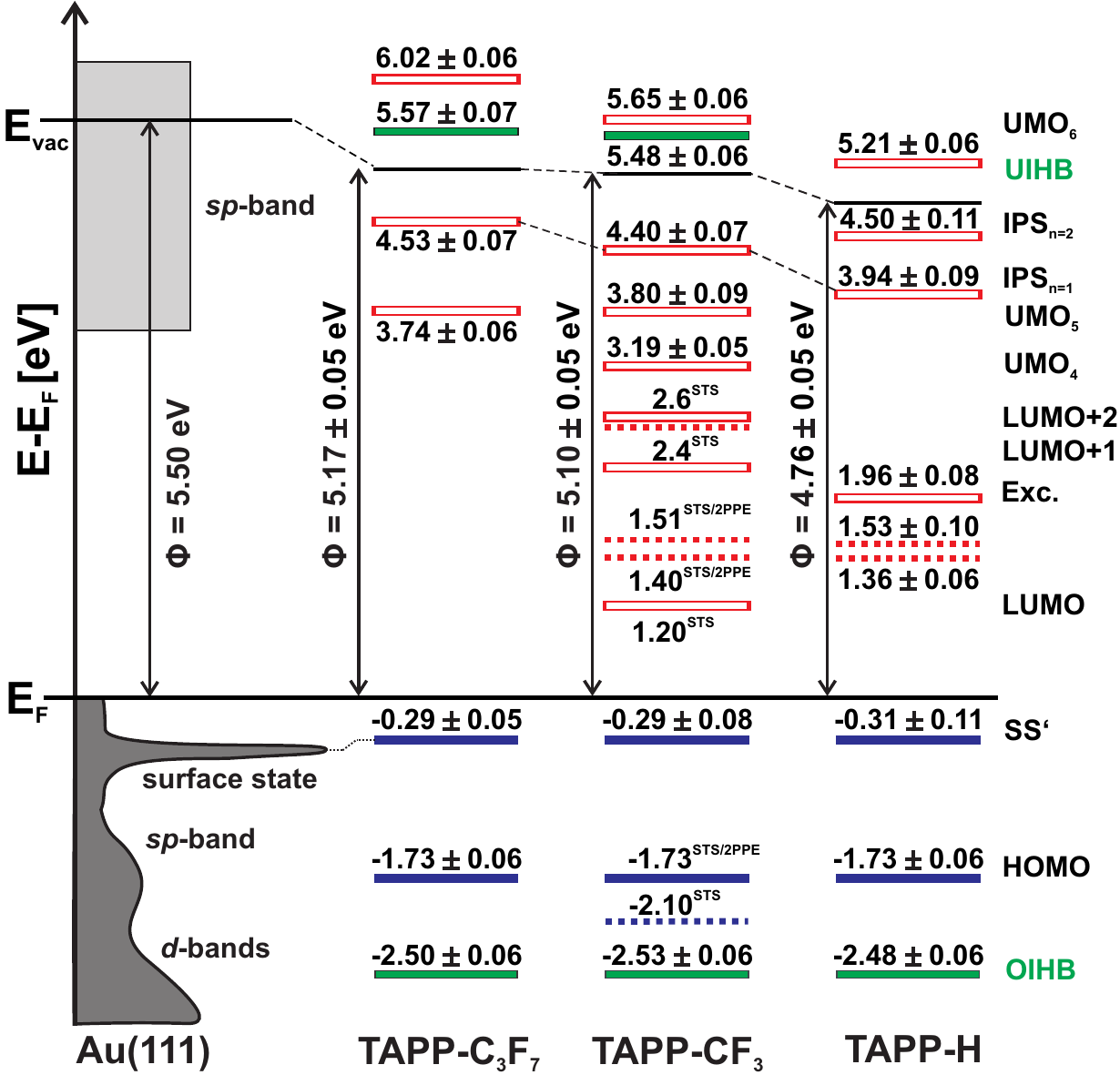}}
\caption{Energy level diagram of 1 ML TAPP-R (R = --H, --CF$_{3}$, --C$_{3}$F$_{7}$) adsorbed on Au(111). $E_F$ denotes the Au(111) Fermi level and $\Phi$ the work function. UMO refers to an unoccupied molecular orbital, OIHB denotes an occupied interfacial hybrid band and UIHB names an unoccupied interfacial hybrid band. $SS^{\prime}$ refers to the shifted surface state and IPS denotes an image potential state. Dashed lines indicate vibronic transitions. Exc. names an excitonic state.}
\label{energydiagram}
\end{figure}

Comparison of the three derivatives allows us to find the following similarities and differences: Adsorption of a monolayer of all these molecules leads to a  shift of the Au(111) surface state ($SS^{\prime}$) of around 200 meV toward the Fermi level. Similar shifts have been observed for many organic molecules on Au(111), where charge transfer to the substrate is negligible, and ascribed to a modification of the image charge and work function ($\Phi$) \cite{gerbert2017molecular, Torrente2008, Scheybal2009, ziroff2009adsorption, yamane2018photoelectron, park2000modification, Stein2019, Stein2017}. Indeed, we find a reduction of the work function for all three molecules, the largest one for TAPP-H by 0.74 eV.

While the work functions for 1ML TAPP-CF$_{3}$/Au(111) ($\Phi$ = 5.10 $\pm$ 0.05 eV) and  1ML TAPP-C$_{3}$F$_{7}$
($\Phi$ = 5.17 $\pm$ 0.05 eV) are very similar the work function possesses a smaller value for 1 ML TAPP-H/Au(111) ($\Phi$ = 4.76 $\pm$ 0.05 eV). STM results indicate an almost equal unit cell area for TAPP-CF$_{3}$ and TAPP-C$_{3}$F$_{7}$ (see Fig. \ref{STMtopo}c and supporting information) thus a similar $\Phi$ value is expected, when considering a similar push back effect.
 In the absence of the fluorinated alkyl side-chains, i.e., for TAPP-H the packing density is most likely increased, leading to a stronger reduction of $\Phi$. Since the image potential states (IPS) are pinned to the vacuum level, their energetic position follows the work function shift.

The HOMO of all three species is located at -1.73 eV, while the unoccupied states vary between the molecules. In the case of TAPP-H/Au(111) we observe vibronic contributions belonging to the LUMO (at 1.36 and 1.53 eV) similar to  TAPP-CF$_{3}$/Au(111). In addition, an unoccupied state at 1.96 eV is observed, which may be attributed to an excitonic state related to the LUMO+1, observed by STS in TAPP-CF$_{3}$/Au(111) at 2.4 eV. Some states in TAPP-H/Au(111), e.g. the UMO$_{4}$ and UMO$_{5}$, and in TAPP-C$_{3}$F$_{7}$/Au(111) the vibronic states associated with the LUMO are not detected. The reasons are not obvious. However in all TAPP-R/Au(111) systems we find interfacial occupied and unoccupied hybrid bands (UIHB and OIHB), which show pronounced dispersions as we will demonstrate by angle-resolved 2PPE in the next section.

\subsection{Band formation at the TAPP/Au(111) interfaces}
\begin{figure}
\centering
\resizebox{0.8\hsize}{!}{\includegraphics{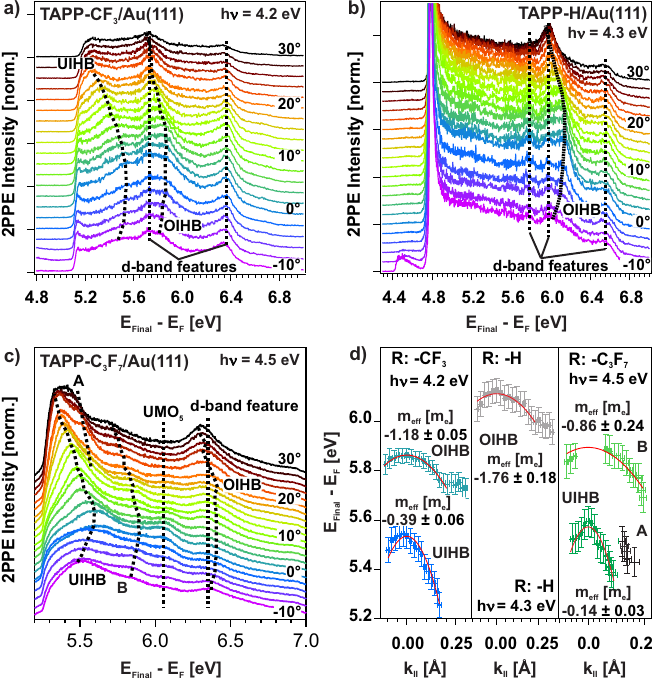}}
\caption{a-c) Angle-resolved 2PPE measurements at the TAPP-R/Au(111) interfaces. d) Hole-like dispersion of the OIHBs, UIHBs, and a state labeled as B (see text) around the $\Gamma$-point ($k_{\|}$ = 0). Parabolic fits yield the effective masses ($m^{*}$) of the states around the $\Gamma$-point.}
\label{dispersion}
\end{figure}

Recently, we observed the dispersion of the UIHB and OIHB for the TAPP-CF$_{3}$ derivative on Au(111) \cite{Stein2021}. Since a different packing or ordering behaviour is expected for the three TAPP derivatives due to the varying size of the side-groups, we studied the influence of the side-chains with respect to interfacial band  formation of the other two TAPP derivatives, namely TAPP-H and TAPP-C$_{3}$F$_{7}$ (see Fig. \ref{molecules}). To be able to conclude that band formation has occurred and to measure dispersion in angle-resolved photoemission, lateral extended interfacial hybrid states (bands) must to be formed. This is only feasible if the molecules
on the surface display well-defined long-range order. To check for similar interfacial bands, we recorded angle-resolved 2PPE spectra also for the two other derivatives (Fig.\,\ref{dispersion}a-c). A compilation of the dispersions is shown in Fig.\,\ref{dispersion}d, evidencing a hole-like dispersion of both the OIHBs and UIHBs.

As already demonstrated in our previous study for the TAPP-CF$_{3}$/Au(111) interface, the effective masses ($m^{*}$) of the OIHB and UIHB  are $m^{*}$ = -1.18 $\pm$ 0.05 $m_{e}$ and $m^{*}$ = -0.39 $\pm$ 0.06 $m_{e}$, respectively \cite{Stein2021}. For the TAPP-H/Au(111) interface the OIHB possesses an effective mass of $m^{*}$ = -1.76 $\pm$ 0.18 $m_{e}$. In the case of the TAPP-C$_{3}$F$_{7}$/Au(111) system the UIHB exhibits an effective mass of $m^{*}$ = -0.14 $\pm$ 0.03 $m_{e}$. The dispersion of the OIHB is visible, but the peak positions cannot be precisely identified, since the 2PPE peak is merged into the broad contribution of the nearby $d$-bands. Thus, a determination of $m^{*}$ is impossible. Additionally, we observe two dispersing states labeled as A and B (Fig.\,\ref{dispersion}c). Unfortunately, these peaks overlap with other 2PPE features, in particular under normal emission ($k_{\|}$ = 0). This renders an assignment to occupied or unoccupied states  \textit{via} photon-energy dependent measurements difficult. In the case of state B, extrapolating an energy $E_{Final}-E_{F}$ = 5.9 eV from the parabolic fit at $k_{\|}$ = 0 would result in an energetic position of an unoccupied state at 1.4 eV or for an occupied state at - 3.1 eV with respect to $E_{F}$. This state has an effective mass of $m^{*}$ = -0.86 $\pm$ 0.24 $m_{e}$.

While hybridization between molecular and metal states occurs in many organic/metal systems, experimental evidence for interfacial band formation via dispersing electronic states is rarely observed. Only for strong electron acceptors (F4TCNQ/Au(111) \cite{Yamane2017, Gerbert2018}, PTCDA/ Ag(110) \cite{Wiessner2013}, NTCDA/Ag(110) \cite{Wiessner2013a}) or a donor (TTF/Au(111) \cite{Gerbert2018}), dispersing interface bands have been proposed. In all cases a charge transfer between molecular states and the surface state of the metallic substrate appears (Fermi-level pinning). In contrast, for the TAPP derivatives adsorbed on Au(111) no charge transfer occurs, since the HOMO is far below $E_F$ (-1.73 eV) for all three adsorbate/substate-systems and the LUMO of TAPP-CF$_{3}$/Au(111) (1.20 eV) as well as the UMO$_{1}$ (1.36 eV) of TAPP-H/Au(111) are energetically far above $E_F$. As discussed in detail for TAPP-CF$_{3}$/Au(111) \cite{Stein2021}, interfacial band formation is only possible if wave function mixing of localized molecular states with delocalized metal bands takes place. The OIHBs are energetically located in the Au(111) d-band region  \cite{Eckardt1984, Courths1986, Takeuchi1991}. Thus, we suggest that occupied TAPP states hybridize with the d-band, leading to the emergence of the OIHBs. The UIHBs lie within the energy regime of the unoccupied metal $sp$ band \cite{Eckardt1984, Takeuchi1991}. Hence, we propose a hybridization between the unoccupied molecular state and this band. Note that due to the planar adsorption geometry of the TAPP-R molecules on the Au(111) surface in the monolayer regime (see supporting information and Refs. \cite{Maass2016, Stein2021}) intermolecular band formation can be excluded.

\section{Conclusion}
 In summary, we have investigated the electronic structure of three tetraazaperopyrene derivatives on Au(111) in great detail by complementary experimental techniques and density functional theory.  We find strong similarities between the derivatives, suggesting common adsorption properties. Most notably all of these derivatives develop interfacial hybrid bands, which we ascribe to hybridization of higher-lying molecular states with the \emph{sp}- and \emph{d}-bands of the substrate. Band formation with the substrate is rather surprising as the energy level alignment of all molecules indicates the absence of charge transfer.
Indeed, we find several indications that the molecules are weakly physisorbed on a Au(111) surface, such as an intact herringbone reconstruction below the molecular layer and the persistence of the surface state. Interfacial band formation while molecular properties are preserved can be considered to be beneficial for charge injection into the organic semiconductor.

\begin{acknowledgement}

We thank N. Krane for discussions and initial code development to calculate vibronic spectra. A.S., D.G., B.G., L.H.G., and P.T. acknowledge funding by the German Research Foundation (DFG) through collaborative research center SFB 1249 "N-Heteropolycylces as Functional Materials" (projects A02 and B06). Funding by the SFB  951 "Hybrid Inorganic/Organic Systems for Opto-Electronics" (Project No. 182087777; K. J. F.) and by the International Max Planck Research School "Functional Interfaces in Physics and Chemistry" (D. R.) is gratefully acknowledged. Funding through the project of the Austrian Science Fund (FWF): Y1157-N36 "MAP-DESIGN" is gratefully acknowledged (A. J., J. J. C., and O. T. H.). Computational results have been achieved using the Vienna Scientific Cluster (VSC)
\end{acknowledgement}

\begin{suppinfo}
Temperature-programmed desorption data for TAPP-R (R: H, CF$_{3}$, C$_{3}$F$_{7}$), calculated molecular interactions of TAPP-CF$_{3}$, two-photon photoemission data for TAPP-H and TAPP-C$_{3}$F$_{7}$, and scanning tunneling microscopy images of TAPP-C$_{3}$F$_{7}$
\end{suppinfo}

\makeatletter
\providecommand{\doi}
  {\begingroup\let\do\@makeother\dospecials
  \catcode`\{=1 \catcode`\}=2 \doi@aux}
\providecommand{\doi@aux}[1]{\endgroup\texttt{#1}}
\makeatother
\providecommand*\mcitethebibliography{\thebibliography}
\csname @ifundefined\endcsname{endmcitethebibliography}
  {\let\endmcitethebibliography\endthebibliography}{}




\end{document}